\documentclass[a4paper,12pt]{article}
\usepackage{amssymb,amsmath}
\usepackage{graphics}
\def\beq{\begin{equation}}
\def\eeq{\end{equation}}
\def\bea{\begin{eqnarray}}
\def\eea{\end{eqnarray}}
\def\nn{\nonumber}

\def\qb#1{\bigg[ #1 \bigg]}

\def\s#1{\bigg\lgroup #1 \bigg\rgroup}

\makeatletter
  \def\@cite#1#2{${\mbox{#1\if@tempswa , #2\fi}}$}
\makeatother

\makeatletter
  \def\@biblabel#1{$^{\mbox{#1}}$}
\makeatother
\renewcommand{\theequation}{\arabic{section}.\arabic{equation}}

\begin{document}
%
%
%
%
\thispagestyle{empty}
\vspace*{3cm}
\begin{center}
{\LARGE\sf{
Nonextensive statistics of relativistic ideal gas}}\\
\end{center}
\begin{center}
{R. Chakrabarti, R. Chandrashekar and S.S. Naina Mohammed}
\end{center}
\begin{center}
Department of Theoretical Physics \\
University of Madras, Guindy Campus \\
Chennai - 600 025, India.
\end{center}
\vfill
\begin{abstract}
We obtain the specific heat in the third constraint scenario for a canonical 
ensemble of a nonextensive extreme relativistic ideal gas in a closed form. 
The canonical ensemble of $N$ particles in $D$ dimensions is well-defined for 
the choice of the deformation parameter in the range 
$ 0 < q < 1 + \frac{1}{D N}$. 
For a relativistic ideal gas with particles of arbitrary mass a perturbative 
scheme in the nonextensivity parameter $(1 - q)$ is developed by employing 
an infinite product expansion of the $q$-exponential, and a \textit {direct 
transformation} of the internal energy from the second to the third constraint 
picture. All thermodynamic quantities may be uniformly evaluated to any 
desired perturbative order. 
\end{abstract}
PACS Number(s): 05.20.-y, 05.70.-a \\
Keywords: Nonextensivity; relativistic ideal gas; perturbative method.

\setcounter{page}{1}
%
%
%
\setcounter{equation}{0}
\section{Introduction}
\label{Intro}

\par
Tsallis [\cite{CT1}] proposed nonextensive statistical mechanics by
 generalizing the functional form of the Boltzmann-Gibbs entropy as
\beq
S_{q}= k \qb{\frac{W^{1-q}-1}{1-q}} \equiv k \ln_{q} W,  \qquad    q \in 
{\mathbb{R}_{+}},
\label{i1}
\eeq 
where the deformation parameter $q$ is taken to be a real positive number as 
this ensures [\cite{SA2}] the stability of the Tsallis entropy. In (\ref{i1}) 
the quantity $k$ is the generalized Boltzmann constant, and $W$ denotes the 
weight. The entropy (\ref{i1}) satisfy a nonlinear, inhomogeneous relation 
\beq
S_q(A+B) = S_q(A)+S_q(B) + (1-q) S_q(A)S_q(B),
\label{i2}
\eeq
where $A$ and $B$ refer to statistically independent systems. The 
nonextensivity 
of the entropy manifest in (\ref{i2}) is governed by the parameter $(1 - q)$. 
The Boltzmann-Gibbs statistics is recovered in the $q \rightarrow 1$ limit. 

\par

The nonextensive statistical mechanics has found wide-ranging applications in 
studies of the systems exhibiting long range interactions [\cite{ND}], long 
time microscopic memory effects [\cite{RT}], anomalous diffusion [\cite{TDF}], 
nonequilibrium phenomena [\cite{AKRG}] and so on. For instance, the formation 
of a new hadronic state of matter known as the quark-gluon plasma that occurs 
in the early stage of the relativistic hadronic collisions exemplifies long 
range interactions as well as long time memory effects [\cite{ALQ}], and, 
consequently, the nonextensive statistical mechanics is expected to be more 
appropriate there than the classical Boltzmann-Gibbs statistics. The rapidity 
spectrum obtained  by using the 
Tsallis distribution is found [\cite{L}] to be in good agreement  with
the experimental data. The data on the distribution of transverse 
momentum of hadrons as well as the differential cross sections in high 
energy $e^{+}\; e^{-}$ collisions bear close resemblance with the theoretical 
analysis [\cite{BCM}, \cite{B}] based on the Tsallis nonextensive statistical 
mechanics. In the context of many body systems endowed with self-gravitating 
long range interactions it has been observed [\cite{J07}] that the power law 
distributions may be achieved using a $q$-kinetic theory based on the
Tsallis statistics.   
As a groundwork for complete theoretical understanding of the relativistic 
heavy ion collisions in the high energy physics regime, and also for 
possible applications in astrophysics, it is imperative to 
study the relativistic ideal gas in the context of nonextensive statistical 
mechanics. This investigation has been initiated in [\cite{AK}]. These authors 
observe that the grand canonical partition function exhibits an essential 
singularity for $q>1$ region, and, consequently, they claim that, in the said 
region, the nonextensive relativistic ideal gas does not exist. 
Making a slight 
departure from the arguments in [\cite{AK}], we, in the present work, consider 
a {\it canonical} ensemble of a fixed number of $N$ molecules of mass $m$ of 
a relativistic ideal
 gas subject to the Tsallis statistics. We produce an exact evaluation of the 
canonical specific heat in the extreme relativistic case ($m \rightarrow 0$).
In $D$ dimensions the generalized partition 
function is nonsingular in the region $0 < q < 1 + \frac{1}{D N}$. For 
an arbitrary mass $m$ we obtain the generalized partition function 
[\cite{TMP}], and the thermodynamic quantities in the second and the third 
constraint pictures as  perturbative series in the nonextensivity parameter 
$(1 - q)$. Towards this purpose, we, as a calculational tool, disentangle the  
$q$-exponential (\ref{i3}) employing a technique developed in [\cite{QC}], and 
previously used [\cite{RCN}] to obtain thermodynamic quantities of a 
nonrelativistic ideal gas obeying the Tsallis statistics. In addition, we 
employ a direct transformation linking the internal energies in the second and 
the third constraint pictures that allows us to evaluate the thermodynamic 
quantities uniformly to any arbitrary prescribed perturbative order in 
$(1 - q)$. For the sake of simplicity we produce them till the 
order $(1-q)^{2}$. 

\par 

The plan of this article is as follows. The extreme or ultra relativistic
gas is discussed in Sec. \ref{ultra}. This is followed by our perturbative 
evaluation  of the thermodynamic quantities for the general case of a 
relativisitic ideal gas in Sec. \ref{Rgpt}. We conclude in Sec. \ref{remark}.
%
%
%
%
%
%
%
%
%
\setcounter{equation}{0}
\section{Ultra relativistic ideal gas}
\label{ultra}
The Hamiltonian of a relativistic ideal gas with particles possessing
$D$-dimensional momenta $\mathbf {p}_{i}\; (i = 1, \cdots, N)$ reads 
\beq
H(p) =  \displaystyle \sum_{i=1}^{N} mc^{2} \left(\sqrt{1 +
\left(\frac{p_i}{mc} 
\right)^{2}} - 1\right), \qquad p_{i} = |\mathbf{p}_{i}|,
\label{ur1}  
\eeq
where $c$ is the velocity of light. In the extreme relativistic case
 $m << p_{i}$\, the Hamiltonian (\ref{ur1}) reduces to
\beq
H(p) = c \displaystyle \sum_{i=1}^{N} p_{i}.
\label{ur2}
\eeq
The generalized partition function in the third constraint [\cite{TMP}] 
approach is given by
\beq
{\overline{\mathcal Z}}_{q}^{(3)}(\beta,V,N) = \frac{1}{N!~ h^{DN}} ~ 
\int d^{DN} x ~ d^{DN} p ~ \exp_{q}\left(- \beta ~ \frac{H(p) - U_{q}^{(3)}}
{\mathfrak{c}^{(3)}}\right), \;\;\; \beta = \frac{1}{k T},
\label{ur3}
\eeq
where $h$ refers to the elementary cell of the one-dimensional phase space, 
and the deformed $q$-exponential is the inverse of the 
$q$-logarithm introduced in (\ref{i1}):
\beq
\exp_{q} (z) = (1+(1-q)z)^{\frac{1}{1-q}}.
\label{i3}
\eeq 
The series expansion for the deformed exponential reads [\cite{JS}]
\beq
\exp_{q} (z) = \displaystyle \sum_{n=0}^{\infty} \frac{z^n}{[n]_{q}!}, 
\qquad [n]_{q} = \frac{n}{1-(1-q)(n-1)}.
\label{i4}
\eeq
We briefly remark here that the following $q$-derivative  
\beq
D_q(x) = \frac{1}{(1-(1-q)(x\frac{d}{dx}))}\frac{d}{dx}
\label{q_der}
\eeq
acts on the monomials $x^{n}$ to produce the $q$-number $[n]_{q}$ defined in 
(\ref{i4}):
\beq
D_q(x) x^n = [n]_q x^{n-1}.
\label{qder_mon}
\eeq
Consequently, the $q$-exponentials represented by the infinite series 
(\ref{i4}) are the eigenfunctions of the 
$q$-derivative introduced in (\ref{q_der}):   
\beq
D_q(x) \exp_q(\alpha x)= \alpha \exp_q(\alpha x).
\label{D_eigen}
\eeq
Incidentally, other $q$-derivatives were introduced in Ref.[\cite{B04}].

\par

\begin{figure} 
\begin{center}
\resizebox{120mm}{!}{\includegraphics{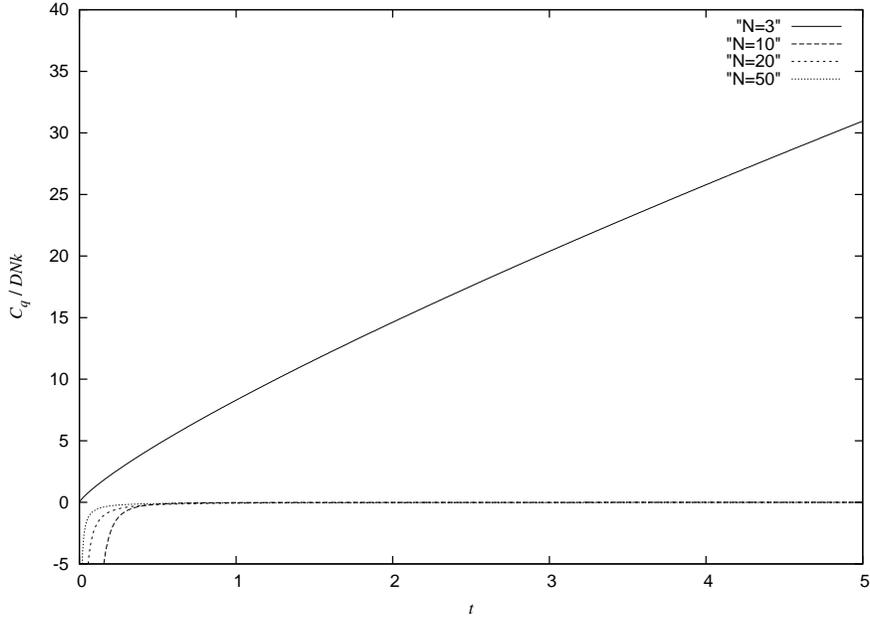}}
\caption{Temperature dependence of specific heat for fixed $q=0.95$ and 
various $N$}
\label{urgN}
\end{center}
\end{figure}

\begin{figure} 
\begin{center}
\resizebox{120mm}{!}{\includegraphics{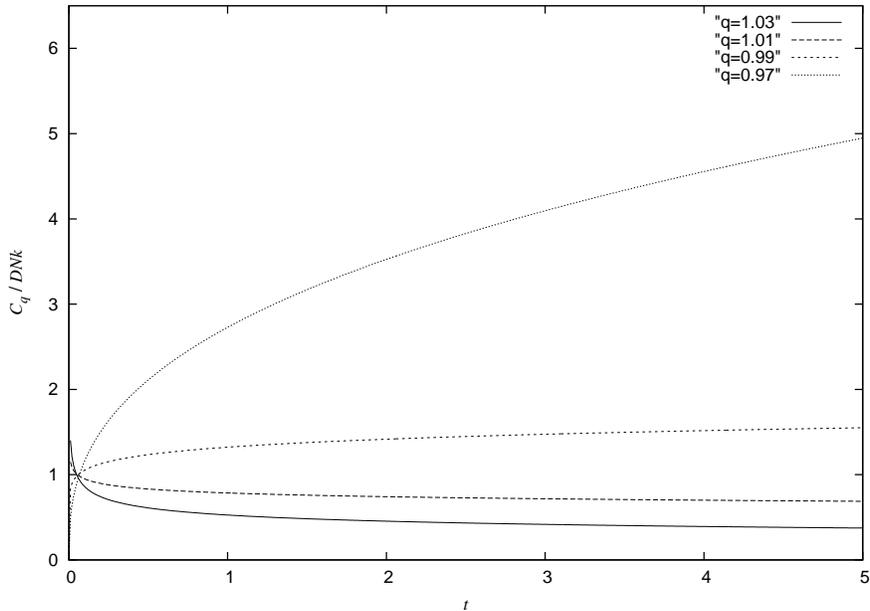}}
\caption{Temperature dependence of specific heat for fixed $N (=3)$ and 
various $q$}
\label{urgq}
\end{center}
\end{figure}

The ensemble probability for an arbitrary energy $E_{j}$ in the third 
constraint framework 
\beq
\mathfrak{p}^{(3)}_{j} (\beta,V,N)  = 
\frac{1}{\overline{\mathcal{Z}}_{q}^{(3)}} \exp_{q}\left(-\beta
\frac{E_{j} - U_{q}^{(3)}}{\mathfrak{c}^{(3)}}\right)
\label{prob_3}
\eeq
leads [\cite{TMP}] to the following sum of the $q$-weights, referred to in 
(\ref{ur3}):  
\beq
\mathfrak{c}^{(3)} \equiv \sum_{j} \big(\mathfrak{p}_{j}^{(3)} \,
(\beta, V, N)\big)^{q} = \left(\overline{\mathcal{Z}}_{q}^{(3)} \right)^{1-q}. 
\label{ur4}
\eeq
For the extreme relativistic Hamiltonian (\ref{ur2}) the generalized partition 
function (\ref{ur3}) reads
\beq
{\overline{\mathcal Z}}_{q}^{(3)}(\beta,V,N) = \mathcal{G} \; 
Z_{m=0}(\beta,V,N) \; \Big(1+(1-q) \frac{\beta}{\mathfrak{c^{(3)}}} \;
U_{q}^{(3)}\Big)^{\frac{1}{1-q}\,+ D N}\;\left(\mathfrak{c}^{(3)}
\right)^{D N},
\label{ur5}
\eeq
where $Z_{m=0}(\beta,V,N) $ is the classical partition function for the ultra 
relativistic gas in arbitrary dimension $D$ 
\beq
Z_{m=0}(\beta,V,N) = \mathcal{W} {\beta}^{-DN},
\label{cl_m0_Z}
\eeq
and the parameters $\mathcal{W}$ 
and $\mathcal{G}$ are given by 
\beq
\mathcal{W} =  \frac{1}{N!}\,\left(\frac{2 \,V\, \pi^{\frac{D}{2}}\,\Gamma(D)}
{(c\, h)^{D}\, 
\Gamma\big(\frac{D}{2}\big)}\right)^{N}, \qquad
\mathcal{G} =  \frac{\Gamma \big(\frac{1}{1-q}+1 \big)}{(1-q)^{DN} 
\Gamma\big(\frac{1}
{1-q} + D N + 1\big)}. \nonumber 
\eeq
The generalized partition function (\ref{ur5}) has simple poles for the values 
of the deformation parameter $q = 1 + \frac{1}{n}, \;
n = 1,..., D N$. The number of singularities equals the number of degrees of 
freedom of the system. The generalized partition function 
is, therefore, well defined in the interval $0 < q < 1+ \frac{1}{D N}$. As the 
number of particles $N$ increases, we observe that the 
poles accumulate towards $q = 1$, the limiting value where statistical 
mechanics becomes extensive.
The internal energy is defined [\cite{TMP}] via the escort probability as
\beq
U^{(3)}_{q} (\beta,V,N) = \big(\mathfrak{c}^{(3)}\big)^{-1} \displaystyle 
\sum_{j} (\mathfrak{p}_{j}^{(3)}(\beta))^q \;E_{j}. 
\label{intgy_3}
\eeq
Converting the above sum to the phase space integration {\it \`{a} la} 
(\ref{ur3}) and employing (\ref{ur4}, \ref{ur5}) we now get 
\beq
 U_q^{(3)}(\beta, V, N) = D N\,\frac{\mathfrak{c}^{(3)}}{\beta}.
\label{i6}
\eeq
Applications of the relations (\ref{ur4}-\ref{intgy_3}) produce the 
explicit solution for the sum of the $q$-weights:
\beq
\mathfrak{c}^{(3)} = (\mathcal{W}\, \mathcal{G})^{\frac{1-q}{1 - (1-q) D N}}\; 
(1 + (1-q) DN )^{\frac{1 + (1-q) D N}{1 - (1-q) D N}}\; 
{\beta}^{-\frac{(1-q) D N}{1 - (1-q) D N}}. 
\label{i7}
\eeq
Above equation in conjunction with (\ref{i6}) now produces the internal 
energy as 
\beq
U^{(3)}_{q}(\beta, V, N) = D N\; (\mathcal{W \,G})^{\frac{1-q}
{1 - (1-q) D N}}\; 
(1 + (1-q) D N)^{\frac{1 + (1-q) D N}
{1 - (1-q) D N}}\; \beta^{-\frac{1}{1 - (1-q) D N}}.
\label{i8}
\eeq 
The specific heat defined as 
\beq
C_{q}^{(3)} \equiv \frac{\partial U_{q}^{(3)}}{\partial T}
\label{C3_def}
\eeq
 reads 
\beq
\frac{C_{q}^{(3)}}{D N k} =  \frac{1}{1 - (1-q) D N}\; 
(\mathcal{W \,G})^{\frac{1-q}{1 - (1-q) D N}}\; 
(1 + (1-q) D N)^{\frac{1 + (1-q) D N}{1 - (1-q) D N}}\; 
{\beta}^{- \frac{(1-q) D N}{1 - (1-q) D N}}.
\label{i9}
\eeq
The extensive limit of the specific heat is readily obtained: 
$C_{q}^{(3)}\Big{|}_{q \rightarrow 1} = D N k$. 
The behaviour of the specific heat with respect to the dimensionless 
scaled temperature 
\beq
t = \left(\frac{V^{1/D}}{ch}\;k T\right)^{\frac{(1-q) D N}{1 - (1-q) D N}}  
\label{scl_temp}
\eeq
for various values of $N$ and $q$ values are shown in the Figs. (\ref{urgN}) 
and (\ref{urgq}), respectively.  
Following [\cite{SA}] we now obtain the specific heat at the physically 
important thermodynamic limit $N \rightarrow \infty, 
V \rightarrow \infty, \rho = \frac{N}{V} \rightarrow $ finite:
\beq
C_{q}^{(3)} \longrightarrow - \frac {c\, h}{\sqrt {\pi}\;(1 - q)}\;
\exp \left(- \Big(1 + \frac{1}{D}\Big)\right)\;
\left(\frac{\Gamma \big(\frac{D}{2}\big)}{2\,\Gamma (D)}\right)^{\frac{1}{D}}\;
\frac{\rho^{\frac{1}{D}}}{T},  
\label{ur11}
\eeq
which implies that the classical $q \rightarrow 1$ limit and the thermodynamic 
limit do not commute. As remarked in [\cite{SA}] the 
$N$-independent negative specific heat realized in (\ref{ur11}) for the extreme
relativistic perfect gas may be of consequence
in astrophysical problems. 
\par
We have also calculated the specific heat in the third constraint picture 
using a detour [\cite{TMP}] via the second constraint framework. 
The results obtained 
through the direct evaluation of the generalized partition function and the 
internal energy transformation method are identical. Further discussions will 
appear in Sec. \ref{Rgpt}.
%
%
%
%
%
%
%
\setcounter{equation}{0}
\section{Relativistic ideal gas: molecules of arbitrary mass}
\label{Rgpt}
The relativistic ideal gas containing particles of arbitrary mass $m$, and 
described by the Hamiltonian (\ref{ur1}) has possible applications in heavy 
ion collisions [\cite{ALQ}-\cite{B}] in nuclear physics. Moreover, in the 
context of self-gravitating systems in astrophysics they are relevant 
[\cite{J07}]. In the opposite limits
$m >> p_{i}$ and $m << p_{i}$ the general relativistic perfect gas reduces to 
the nonrelativistic case and the extreme relativistic case, respectively.
For the relativistic ideal gas we obtain the thermodynamic quantities 
perturbatively by treating $(1 - q)$ as the series parameter. In 
the present section we employ the second constraint approach [\cite{TMP}] as 
an intermediate step for evaluating the thermodynamic quantities. The physical 
variables obtained in the second constraint method are transformed 
[\cite{TMP}] to the respective quantities in the third constraint framework by 
introducing a fictitious temperature $\beta^{\prime}$ that provides for the 
correspondence between the alternate constraints. The ensemble probability and 
the partition function in the second constraint [\cite{CT2}] approach read
\beq
\mathfrak{p}_{j}^{(2)} (\beta,V,N)  = \frac{\exp_{q} (- \beta\, E_{j})}
{\mathcal{Z}_{q}^{(2)}}, \qquad 
\mathcal{Z}_{q}^{(2)} = \sum_{j}\,\exp_{q}(- \beta\, E_{j}).
\label{prob_2}
\eeq
The above probability is linked [\cite{TMP}] to that of the third constraint 
approach as
\beq 
\mathfrak{p}_{j}^{(3)} (\beta) = \mathfrak{p}_{j}^{(2)} (\beta^{\prime}),
\label{pro_link}
\eeq
where the general transformation rule for the temperature reads
\beq
\beta = \beta^{\prime}\;\frac{\mathfrak{c}^{(2)} (\beta^{\prime})}{1 - (1 - q) 
\beta^{\prime}\,\frac{ U_{q}^{(2)} (\beta^{\prime})}
{\mathfrak{c}^{(2)} (\beta^{\prime})}},\qquad \mathfrak{c}^{(2)} (\beta) = 
\sum_{j} \Big(\mathfrak{p}_{j}^{(2)} (\beta)\Big)^{q}.
\label{gen_tran}
\eeq
In the above equation the internal energy in the second constraint approach is 
defined [\cite{CT2}] as 
\beq
U_{q}^{(2)}(\beta) = \displaystyle \sum_{j} \big(\mathfrak{p}_{j}^{(2)} 
(\beta)\big)^{q}\; E_{j}.
\label{int_en_2}
\eeq
Converting the sum over states in (\ref{prob_2}) to an integration over the
phase space for the Hamiltonian (\ref{ur1}) the partition function in the 
second constraint approach may be recast  as
\beq
{\mathcal Z}_{q}^{(2)}(\beta, V, N) = \frac{1}{N!~ h^{D N}} ~ \int d^{D N} x ~ 
d^{D N} p ~ \exp_{q} (- \beta ~ H(p)).
\label{cpr2}
\eeq
The integral over the phase-space in (\ref{cpr2}) may be performed exactly in 
the extreme relativistic case in the $m \rightarrow 0$ limit: 
\beq
\mathcal{Z}^{(2)}(\beta,V,N)\Big{|}_{m=0} = Z_{m=0} (\beta,V,N) 
\left(\frac{\Gamma(\frac{2-q}{1-q})}{(1-q)^{DN} \Gamma(\frac{2-q}{1-q}+DN)} 
\right),
\label{Z2_m0}
\eeq
where the classical partition function $Z_{m=0} (\beta,V,N)$ for the extreme 
relativistic perfect gas is given in (\ref{cl_m0_Z}). In the general case 
with particles of arbitrary mass we proceed towards evaluating the integral 
(\ref{cpr2}) perturbatively.
Assuming the partition function (\ref{cpr2}) to be well behaved in the
 neighborhood of $q = 1$, we  follow [\cite{QC},\cite{RCN}] to 
disentangle the $q$-exponential (\ref{i3}) as an infinite product series of 
ordinary exponentials: 
\beq
\exp_{q} (- \beta ~ H(p)) = \exp \left( - \displaystyle \sum_{n=1}^{\infty} 
\frac{(1-q)^{n-1}}{n} {\beta}^{n} H^{n}(p)\right) 
\equiv \widehat{\mathcal{D}}(d_{\beta})~\exp (- \beta ~ H(p)).
\label{cpr3}
\eeq
In (\ref{cpr3}) the operator valued series 
$\widehat{\mathcal{D}}(d_{\beta})$ reads
\bea
\widehat{\mathcal{D}}(d_{\beta}) &=& 1 - \frac{(1-q)}{2}\;d^{(2)}_{\beta} + 
\frac{(1-q)^2}{3} \Big(d^{(3)}_{\beta}+ \frac{3}{8}\,
d^{(4)}_{\beta}\Big) + \ldots ,
\label{cpr4}
\eea
where $d^{(n)}_{\beta} = {\beta}^n \frac{{\partial}^n}{\partial {\beta}^n}$.  
The operator (\ref{cpr4}) links the partition function 
(\ref{cpr2}) in the second constraint approach with the classical 
Boltzmann-Gibbs partition function $Z (\beta, V, N)$ as   
\beq
{\mathcal Z}_{q}^{(2)}(\beta,V,N) = \widehat{\mathcal{D}}(d_{\beta})\;
Z (\beta, V, N),
\qquad Z (\beta, V, N) = \frac{1}{N!} (Z (\beta, V, 1))^{N}.
\label{part_rel}
\eeq
Henceforth, for the purpose of simplicity, we consider $D = 3$.
The phase-space integral of the classical partition function for a single 
particle reads 
\beq
Z (\beta, V, 1) = \frac{1}{h^{3}}\;\int d^{3}x\, d^3p\, 
\exp \left(- \beta\,mc^{2} \left(\sqrt{1 + \Big(\frac{p}{mc}\Big)^{2}} 
- 1\right)\right), 
\label{Z_1_int}
\eeq
and may be expressed [\cite{GNS95}] via the modified Bessel function of the 
second kind 
$K_{n}(z),\;n = 2$:
\beq
Z (\beta, V, 1) =\frac{4 \pi V\, \exp(u)\, K_{2}(u)}{\lambda^{3}\, u}, \qquad 
\lambda = \frac{h}{m c},\quad u = \beta m c^{2}.
\label{Z_1_K}
\eeq
For future use we express $Z (\beta, V, N)$ in a factorized form:
\beq
Z (\beta, V, N) = g(V)\;f(u), \quad  g(V) = \left(\frac{4 \pi V}
{{\lambda}^{3}}\right)^{N},\;\;\; f(u) = \frac{1}{N!}\;\left(\frac{\exp(u)\;
K_{2}(u)}{u}\right)^{N}.
\label{Z_factor}
\eeq
The recipe (\ref{part_rel}, \ref{Z_1_K}) now produces the partition function 
in the second constraint approach as an infinite perturbative series in the
nonextensivity parameter $(1 - q)$:
\beq
{\mathcal Z}_{q}^{(2)} (\beta, V, N) = Z(\beta, V, N) \; \left(\displaystyle 
\sum_{n=0}^{\infty} (-1)^{n}\, (1-q)^{n} \sum_{\ell=0}^{2 n} 
\alpha_{n \ell}(u)\; (\mathcal{K} (u))^{\ell}\right),
\label{cpr6}
\eeq
where we have defined 
\beq 
\mathcal{K} (u) \equiv u \; \frac{K_{1}(u)}{K_{2}(u)}.
\label {K_ratio}
\eeq
 The coefficients $\alpha_{n\,\ell}(u)$
for the first few orders are listed below:
\bea
\alpha_{00}(u) &=& 1,\qquad\qquad\qquad \alpha_{10}(u) = \frac{3}{2} 
(1 + 3 N) N - 3 N^{2} u + \frac{1}{2} (1 + N) N u^{2},\nn\\ 
\alpha_{11}(u) &=& - \frac{3}{2} (1 - 2 N) N  -  N^{2}  u,\qquad\qquad 
\alpha_{12}(u) = - \frac{1}{2} (1 - N)\; N,\nn\\
\alpha_{20}(u) &=& \frac{1}{8} (2 + 27 N  + 90 N^{2} + 81 N^{3}) N 
- \frac{9}{2} (1 + 3 N) N^{3} u \nn\\
& &+ \frac{1}{8}(11 - 42 N + 48 N^{2} + 54 N^{3}) N  u^{2} 
+ \frac{1}{6} (15 - 25 N - 9 N^{2}) N^{2} u^{3} \nn\\ 
& &- \frac{1}{8} (2 - 3N - 6 N^{2} - N^{3}) N u^{4},\nn\\
\alpha_{21}(u) &=& - \frac{1}{4} (11 - 45 N + 63 N^{2} - 54 N^{3}) N  
- \frac{3}{2} (4 -10 N + 9 N^{2}) N^{2} u \nn\\ 
& &+ \frac{1}{12} (44 - 93 N + 15 N^{2} + 54 N^{3}) N u^{2} 
+ \frac{1}{2} (2 - 3 N - N^{2}) N^{2} u^{3},\nn \\
 \alpha_{22}(u) &=& - \frac{3}{8} (25 - 63 N + 56 N^{2} - 18 N^{3}) N  
- \frac{1}{2} (11 - 20 N + 9 N^{2}) N^{2}  u \nn\\ 
& & + \frac{1}{4} (4 - 7 N + 3 N^{3}) N u^{2},\nn \\
 \alpha_{23}(u) &=& - \frac{1}{12} (62 - 129 N + 85 N^{2} - 18 N^{3})\; N 
- \frac{1}{2} (2 - 3 N + N^{2})  N^{2}u, \nn\\
 \alpha_{24}(u) &=& - \frac{1}{8} (6 - 11 N  + 6 N^{2} - N^{3})\; N. 
\label{Z2_coeff}
\eea
In the second constraint framework the internal energy (\ref{int_en_2}) may be
recast as
\beq
U_{q}^{(2)}(u) = - m c^{2} \frac{\partial}{\partial u} 
\ln_{q} {\mathcal{Z}}_{q}^{(2)}(u). 
\label{cpr{8}}
\eeq
The expressions (\ref{cpr6}) and (\ref{Z2_coeff}) may now be employed to 
produce the following series expansion for the internal energy
$U_{q}^{(2)}(u)$:
\beq
\frac{U_{q}^{(2)}(u)}{N k T} = (Z(u,V,N))^{1-q} \left(\displaystyle 
\sum_{n=0}^{\infty}
 (1-q)^{n} \sum_{\ell = 0}^{2n+1} 
\rho_{n \ell}(u) \Big({\mathcal{K}}(u)\Big)^{\ell}\right),
\label{int_en_sec}
\eeq
where the first few coefficients read
\bea
\rho_{00}(u) &=& 3-u, \qquad\qquad \rho_{01}(u) = 1,
\qquad \qquad\rho_{10}(u) 
= - 3 N u + \frac{1}{2} (5 - 4 N) u^{2} + N u^{3}, \nn \\
\rho_{11}(u) &=& 6 (2 N - 1) - 5 N u + (1 - N) u^{2},\;\; 
\rho_{12}(u)  =  \frac{14 N - 11}{2} - N u,\;\; \rho_{13}(u) = N-1, \nn\\
\rho_{20}(u) &=& - \frac{9}{2} (1 + 3 N) N + \frac{3}{2} (1 + 9 N) N u 
- \frac{1}{2} (11 - 48 N + 48 N^{2}) u^{2}\nn\\ 
& & - \frac{1}{2} (26 - 41 N) N u^{3} 
+ \frac{1}{6} (28 - 48 N - 9 N^{2}) u^{4} + (1 - N) N u^{5}, \nn\\
\rho_{21}(u) &=& \frac{1}{2} (22 - 102 N + 81 N^{2}) 
                    + \frac{1}{2}(57-102N) N u
               - \frac{1}{4} (163 - 352 N + 138 N^{2}) u^{2} \nn\\ 
& & - 3(6 - 7N) N u^{3} + (2 - 3 N) u^{4}, \nn\\
\rho_{22}(u) &=& \frac{1}{4} (311 - 786 N + 516 N^{2}) 
+ \frac{1}{2} (110-147 N) N u
 - \frac{1}{6} (175 - 312 N + 90 N^{2}) u^{2} \nn\\ 
& & - 4 (1 - N) N u^{3}, \nn\\
\rho_{23}(u) &=& \frac{1}{4} (323 - 664 N + 350 N^{2}) + 3 (8 - 9N) N u 
- (5- 8 N +2 N^{2}) u^{2}, \nn\\
\rho_{24}(u) &=& \frac{5}{2} (11 - 20 N + 9 N^{2}) +3 (1 - N) N u,\quad 
\rho_{25}(u)  = (3 - 5 N + 2 N^{2}).
\label{U2_per}
\eea
In constructing the transformation leading to the thermodynamic quantities 
pertaining to the third constraint approach we first express 
the weight factor in (\ref{gen_tran})  in terms of an integral over the phase 
space, and subsequently use a perturbative approach \textit {\`{a} la} 
(\ref{part_rel}):  
\beq
\mathfrak{c}^{(2)}(\beta) = \frac{1}{(\mathcal{Z}_{q}^{(2)}(\beta, V, N))^{q}} 
\mathcal{\widehat{R}}(d_{\beta})\;Z (\beta, V, N),
\label{weight2_per}
\eeq
where the operator-valued series reads 
\beq
\mathcal{\widehat{R}}(d_{\beta}) = 1 - (1-q)\, \Big(d_{\beta}^{(1)} 
+ \frac{1}{2} d_{\beta}^{(2)}\Big) + (1-q)^{2}\, 
\Big(d_{\beta}^{(2)} + \frac{5}{6} d_{\beta}^{(3)} + \frac{1}{8} 
d_{\beta}^{(4)}\Big) + \ldots.
\label{R-expn}
\eeq
We follow (\ref{weight2_per}) to compute the weight factor 
$\mathfrak{c}^{(2)}(\beta)$, and subsequently substitute it in the 
transformation equation (\ref{gen_tran}). To eliminate a trivial kinematical 
dependence on the volume of the nonextensive system in our evaluation of its
specific heat, we introduce an appropriately scaled variable. Using the 
factorized form of the classical partition function (\ref{Z_factor}) we define 
\beq
\mathfrak{u} = \frac{u}{(g(V))^{1-q}}
\label{vol_scale}
\eeq 
and compute the inverse 
transformation in a series as 
\beq
u^{\prime} = \mathfrak{u} \left(1 + \displaystyle \sum_{n=1}^{\infty} (1-q)^{n}
 \sum_{\ell = 0}^{2n-1} \mathfrak{g}_{n\,\ell}(\mathfrak{u}) 
\Big(\mathcal{K}(\mathfrak{u})\Big)^{\ell}\right),
\label{uprime_u}
\eeq
where the first few perturbative coefficients read 
\bea
\mathfrak{g}_{10}(\mathfrak{u}) &=& - (6 N + \ln f(\mathfrak{u})) 
+ 2 N \mathfrak{u},\qquad \qquad \qquad 
\mathfrak{g}_{11}(\mathfrak{u}) = 2 N , \nn\\
\mathfrak{g}_{20}(\mathfrak{u}) &=& \frac{3}{2} 
(1 + 9 N + 2 \ln f(\mathfrak{u})) N + \frac{1}{2} (\ln f(\mathfrak{u}))^{2} 
- 3 (5 N + \ln f(\mathfrak{u})) N \mathfrak{u} \nn\\ 
& & - \frac{1}{2} (9 + 5 N - 4 \ln f(\mathfrak{u}))
N \mathfrak{u}^{2} + 2 N^2 {\mathfrak{u}}^3,\nn\\
\mathfrak{g}_{21}(\mathfrak{u}) &=& \frac{3}{2} 
(7 + 22 N + 6 \ln f(\mathfrak{u})) N - 13 N^2 {\mathfrak{u}} - 2 (1 + N) N 
{\mathfrak{u}}^{2},  \nn\\
\mathfrak{g}_{22}(\mathfrak{u}) &=& \frac{1}{2}
(21 + 31 N + 4 \ln f(\mathfrak{u})) N - 2 N^2 \mathfrak{u}, \quad  
\mathfrak{g}_{23}(\mathfrak{u}) =  2 (1 + N) N.
\eea

\begin{figure}
\begin{center}
\resizebox{120mm}{!}{\includegraphics{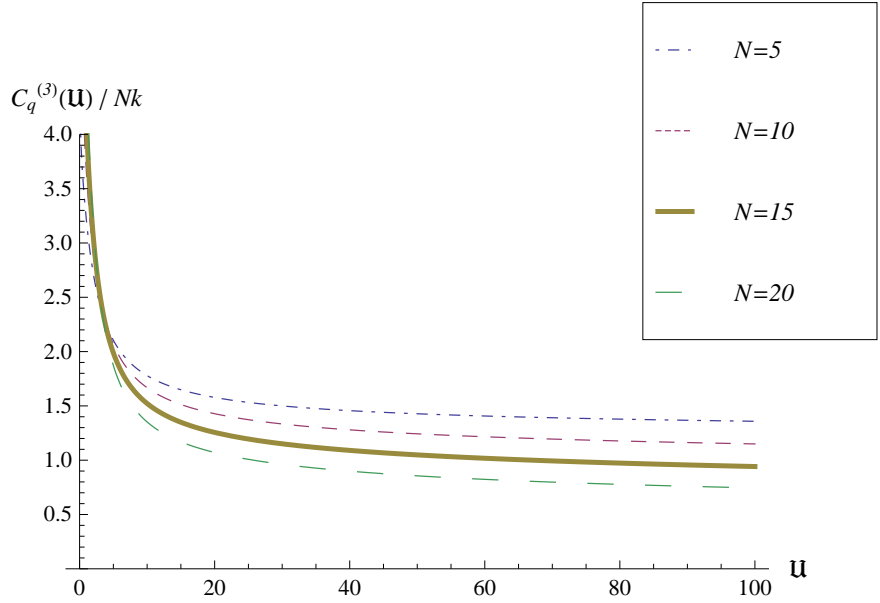}}
\caption{Dependence of specific heat on $\mathfrak{u}$ for fixed $q=0.995$ 
and various $N$}
\label{urgN1}
\end{center}
\end{figure}

\begin{figure}
\begin{center}
\resizebox{120mm}{!}{\includegraphics{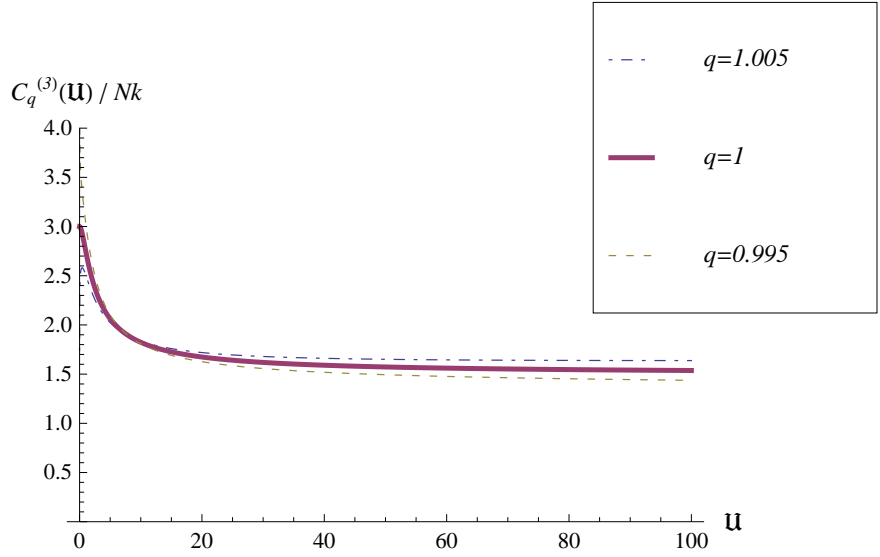}}
\caption{Dependence of specific heat on $\mathfrak{u}$ for fixed $N=3$ 
changing $q$}
\label{egq}
\end{center}
\end{figure}

The internal energies in the second and the third constraint pictures may be 
directly interrelated. Employing the respective definitions (\ref{int_en_2}) 
and (\ref{intgy_3}) in conjunction with the ensemble probabilities 
(\ref{prob_2}) and (\ref{prob_3}) pertaining to these two pictures, the 
internal energy in the third constraint scenario may be expressed in terms of 
the fictitious temperature $\beta^{\prime}$ as  
\bea
U_{q}^{(3)}(\beta) = \frac{U_{q}^{(2)}({\beta}^{\prime})}
{\mathfrak{c}^{(2)}(\beta^{\prime})}.
\label{int_3_trans}
\eea
The above transformation method may be readily extended to any other 
thermodynamic average. The compendium of structures described in 
(\ref{int_en_sec}-\ref{int_3_trans}) now produce the internal energy of an 
arbitrary relativistic gas in the third 
constraint picture as a perturbative series in $(1-q)$: 
\beq
\frac{U^{(3)}_{q}(\mathfrak{u})}{N k T} =  \displaystyle \sum_{n=0}^{\infty} 
(1-q)^{n} \left(\sum_{\ell = 0}^{2 \ell + 1} \varrho_{n \ell}(\mathfrak{u}) 
\Big(\mathcal{K}(\mathfrak{u})\Big)^{\ell}\right).
\label{U3_series}
\eeq
To derive the above expression for the internal energy we have employed 
rescaling of the argument given in Eq. (\ref{bess_sr_u2w}) in the Appendix.
The first few coefficients $\varrho_{n \ell}$ in (\ref{U3_series}) are 
listed below:
\bea
\varrho_{00}(\mathfrak{u}) &=& 3-\mathfrak{u},\qquad \qquad \qquad 
\varrho_{01}(\mathfrak{u}) = 1, \nn \\
\varrho_{10}(\mathfrak{u}) &=& 3 N (3 - \mathfrak{u}) + \frac{1}{2} (5 + 6 N)
\mathfrak{u}^2 - N \mathfrak{u}^3 
+ (3 + \mathfrak{u}^2)\ln f(\mathfrak{u}),\nn\\
\varrho_{11}(\mathfrak{u}) &=& -6 (1 + N) + 3 N \mathfrak{u} 
+ (1 + N) \mathfrak{u}^{2} - 3 \ln f(\mathfrak{u}), \nn \\
\varrho_{12}(\mathfrak{u}) &=&  -\frac{11}{2} - 6 N 
+ N \mathfrak{u} - \ln f(\mathfrak{u}), \qquad  
\varrho_{13}(\mathfrak{u}) = - (1 + N), \nn \\
\varrho_{20}(\mathfrak{u}) &=&  - \frac{9}{2}(1 - 9 N) N - 18 N^2 \mathfrak{u}
- \frac{1}{2} (11 + 57 N - 3 N^2) \mathfrak{u}^2\nn\\ 
& &+ \frac{1}{2} (17 + 12 N) N \mathfrak{u}^3 + \frac{1}{6}(28 + 15 N + 6 N^2) 
\mathfrak{u}^4 - (1 + N) N \mathfrak{u}^5 \nn\\ 
& & + \Big(3 N (6 - \mathfrak{u}) - \frac{1}{2} (17 + 12 N) \mathfrak{u}^2 
+ 4 N \mathfrak{u}^3 + (1 + N) \mathfrak{u}^4\Big) \ln f(\mathfrak{u}) \nn \\
& & + \frac{3}{2} (\ln f(\mathfrak{u}))^2 - 2 \mathfrak{u}^2 
(\ln f(\mathfrak{u}))^2, \nn \\
\varrho_{21}(\mathfrak{u}) &=& 11 + 63 N + \frac{9}{2} N^2
- 6 (3 + 2 N) N \mathfrak{u} \nn\\  
& &- \frac{1}{4} (163 + 214 N + 114 N^2) \mathfrak{u}^{2} + 16 (1 + N) N 
\mathfrak{u}^{3} + (2 + N) \mathfrak{u}^{4}\nn\\ 
& &+ \Big(6 (3 + 2 N) - (16 + 9 N) \mathfrak{u} 
- 16 N \mathfrak{u}^2 + 2 N \mathfrak{u}^3\Big) \ln f(\mathfrak{u})\nn\\
& &+ \frac{1}{2} (9 - 2 \mathfrak{u}^2)\; (\ln f(\mathfrak{u}))^2,\nn\\
\varrho_{22}(\mathfrak{u}) &=& \frac{1}{4} (311 + 600 N + 258 N^2)  
- \frac{1}{2} (89 + 84 N) N \mathfrak{u} - \frac{1}{6} (175 + 162 N + 90 N^2) 
\mathfrak{u}^{2}\nn\\ 
& & + 4 (1 + N) N \mathfrak{u}^{3} + \frac{1}{2} \Big(89 + 84 N  
- 20 N \mathfrak{u} - 8 (1 + N) \mathfrak{u}^2\Big) \ln f(\mathfrak{u})\nn\\ 
& & + 5 (\ln f(\mathfrak{u}))^2, \nn\\ 
\varrho_{23}(\mathfrak{u}) &=& \frac{1}{4} (323 + 430 N + 258 N^2)  
- 22 (1 + N) N \mathfrak{u} - (5 + 4 N + 2 N^{2}) \mathfrak{u}^{2} \nn\\ 
& & + (22 (1 + N)  - 2 N \mathfrak{u}) \ln f(\mathfrak{u}) 
+  (\ln f(\mathfrak{u}))^2, \nn\\ 
\varrho_{24}(\mathfrak{u}) &=& \frac{1}{2} (55 + 61 N + 40 N^2) 
- 3 (1 + N) N \mathfrak{u} + 3 (1 + N) \ln f(\mathfrak{u}),\nn\\
\varrho_{25}(\mathfrak{u}) &=& 3 + 3 N + 2 N^2.
\label{U3_coeffs}
\eea
With the internal energy (\ref{U3_series}, \ref{U3_coeffs}) at hand, the 
definition of the specific heat (\ref{C3_def}) in the third constraint 
scenario may be employed in conjunction with the scaling equation 
(\ref{vol_scale}) for obtaining the following perturbative series in the 
nonextensivity parameter $(1 - q)$:
\bea
\frac{C^{(3)}_{q} (\mathfrak{u})}{N k} = C_{0} (\mathfrak{u}) + (1-q)\; 
C_{1} (\mathfrak{u}) + (1-q)^{2}\; C_{2} (\mathfrak{u}) + \ldots,
\label{C3_expn} 
\eea
where the extensive Boltzmann-Gibbs limit is given by $C_{0}(\mathfrak{u})$.
We now enlist the first few coefficients in the rhs of (\ref{C3_expn}): 
\bea 
C_{0}(\mathfrak{u}) &=& 3 + \mathfrak{u}^2 - 3  \mathcal{K}(\mathfrak{u}) 
- (\mathcal{K}(\mathfrak{u}))^{2},\nn\\
C_{1} (\mathfrak{u}) &=& 3 N (6  - \mathfrak{u}) - \frac{1}{2} (17 + 12 N)
\mathfrak{u}^2+ 4 N \mathfrak{u}^3 + (1 + N) \mathfrak{u}^4 
+ (3 - 4\mathfrak{u}^2) \ln f(\mathfrak{u})\nn\\
& &+ \s{3 \Big(6 + 4 N + 3 \ln f(\mathfrak{u})\Big)  
- 9 N \mathfrak{u} - 2 \Big(8 + 8 N + \ln f(\mathfrak{u})\Big)\mathfrak{u}^{2} 
+ 2 N \mathfrak{u}^{3}} \mathcal{K}(\mathfrak{u})\nn\\ 
& &+ \s{\frac{1}{2} \Big(89 + 84 N + 20 \ln f(\mathfrak{u})\Big) 
- 10 N \mathfrak{u} - 4 (1 + N) \mathfrak{u}^{2}} 
(\mathcal{K}(\mathfrak{u}))^2\nn\\ 
& &+ 2 \s{\Big(11 (1 + N) + \ln f(\mathfrak{u})\Big) 
- N \mathfrak{u}} (\mathcal{K}(\mathfrak{u}))^3 
+ 3 (1 + N)  (\mathcal{K}(\mathfrak{u}))^4,\nn\\
C_2 (\mathfrak{u}) &=& - \frac{9}{2} (1 - 21 N) N - 27 N^2 \mathfrak{u} 
+ \frac{3}{2} (11 + 44 N - 8 N^2 ) \mathfrak{u}^2 
- \frac{1}{2} (53 + 12 N) N \mathfrak{u}^3\nn \\ 
& & -\frac{1}{4} ( 219 + 232 N + 130 N^{2}) \mathfrak{u}^4 
+19 (1 + N) N \mathfrak{u}^{5}+ (2 + N)\mathfrak{u}^6 \nn\\ 
& & + \frac{1}{2}\Big(54 N - 6 N \mathfrak{u} + (53 + 12 N) \mathfrak{u}^2 
- 26 N \mathfrak{u}^3 - 38 (1 + N) \mathfrak{u}^4 + 4 N \mathfrak{u}^5\Big) 
\ln f(\mathfrak{u})\nn\\ 
& & + \frac{1}{2} \Big( 3 + 13 \mathfrak{u}^2 - 2 \mathfrak{u}^4 \Big) 
(\ln f(\mathfrak{u}))^2 -\s{\frac{1}{2} \Big(66 + 270 N - 81 N^2 + 12 (9+N) 
\ln f(\mathfrak{u})\nn\\ 
& & + 27 (\ln f(\mathfrak{u}))^2\Big)  
+ 3 (18 + 2 N + 9 \ln f(\mathfrak{u})) N \mathfrak{u} 
+ \frac{1}{4} \Big(1437 + 2044 N + 906 N^{2}\nn\\ 
& & + 4 (169+154N) \ln f(\mathfrak{u}) + 60 (\ln f(\mathfrak{u}))^2\Big)
\mathfrak{u}^{2} -  (169 + 154 N + 30 \ln f(\mathfrak{u})) N
\mathfrak{u}^{3} \nn\\ 
& & - \frac{1}{3} \Big(217 + 180 N + 93 N^2 + 24 (1 + N) 
\ln f(\mathfrak{u})\Big) 
\mathfrak{u}^{4} + 8 (1 + N)N \mathfrak{u}^{5}} \mathcal{K}(\mathfrak{u})\nn\\
& & -\frac{1}{4}\s{2221 + 1318 \ln f(\mathfrak{u})
+ 158 (\ln f(\mathfrak{u}))^2 + 2 \Big(1923 + 534 
\ln f(\mathfrak{u}) + 636 N \Big) N\nn\\ 
& & - 2(659 + 534 N + 158 \ln f(\mathfrak{u})) N\mathfrak{u}
- 2 \Big(1091 + 236 \ln f(\mathfrak{u})+8(\ln f(\mathfrak{u}))^2 \nn\\ 
& & + 4(1182 + 232 \ln f(\mathfrak{u}) + 678 N) N \Big)\mathfrak{u}^{2}
+ 4(118 + 8 \ln f(\mathfrak{u}) + 116 N) N\mathfrak{u}^{3} \nn\\ 
& & + 4(17 + 13 N + 6 N^2) \mathfrak{u}^{4}}(\mathcal{K}(\mathfrak{u}))^2
-\s{ \frac{1}{4} \Big(4175 + 5488 N + 2922 N^2 \nn \\
& & + (1324 + 1240 N)\ln f(\mathfrak{u}) + 84 (\ln f(\mathfrak{u}))^2 \Big)  
- (331 + 310 N + 42\ln f(\mathfrak{u})) N \mathfrak{u} \nn\\ 
& & -\frac{1}{3} \Big(700 + 672 N + 402 N^2 + 60(1 + N) 
\ln f(\mathfrak{u})\Big) \mathfrak{u}^2 
+20 (1 + N) N \mathfrak{u}^3} (\mathcal{K}(\mathfrak{u}))^{3}\nn\\ 
& & -\s{\frac{1}{4} \Big(2619 + 2996 N + 1850 N^2 +4 (111 + 109 N) 
\ln f(\mathfrak{u})
+ 12 (\ln f(\mathfrak{u}))^2 \Big) \nn\\ 
& & - (111 + 109 N + 6 \ln f(\mathfrak{u})) N \mathfrak{u}
- (30 + 27 N + 16 N^2)
\mathfrak{u}^{2}} (\mathcal{K}(\mathfrak{u}))^{4} \nn\\ 
& & -\s{167 + 176 N + 115 N^2 + 12(1 + N)\ln f(\mathfrak{u})  
- 12 (1 + N) N \mathfrak{u} } (\mathcal{K}(\mathfrak{u}))^{5} \nn\\ 
& & - 5 (3 + 3 N + 2 N^2)(\mathcal{K}(\mathfrak{u}))^{6}.
\label{cprcr}
\eea 
The dependence of the specific heat (\ref{cprcr}) on the dimensionless 
variable $\mathfrak{u}$ for numerous values of $N$ and $q$ are displayed in 
the Figs. (\ref{urgN1}) and (\ref{egq}), respectively.

\par

Above perturbative evaluation of the specific heat hinges on the series 
(\ref{U3_series}) for the internal energy in the third constraint picture, 
and the thermodynamic transformation property (\ref{int_3_trans}) that is 
based on the equivalence of the ensemble probabilities (\ref{pro_link}). 
An alternate determination of the specific heat utilizing the generalized 
partition function (\ref{ur3}) is well-known [\cite{TMP}]. Integration of 
the following thermodynamic relation [\cite{TMP}] 
\beq
{\mathfrak{u}} \frac{\partial U_{q}^{(3)}}{\partial {\mathfrak{u}}} =
\frac{\partial}{\partial {\mathfrak{u}}} \ln_{q} 
{\overline{\mathcal{Z}}}_{q}^{(3)}
\label{U3_Z3}
\eeq
allows one to determine the internal energy in the third constraint picture,
which, in turn, produces the specific heat via the definition 
(\ref{C3_def}). In the present case, an explicit analytical integration of the 
differential equation (\ref{U3_Z3}) even as a perturbative series turns out to 
be difficult. The specific heat, however, may be directly extracted 
[\cite{RCN}] from the generalized partition function as follows:
\beq
C_{q}^{(3)}(\mathfrak{u}) = - k {\mathfrak{u}} \; \frac{\partial}
                {\partial {\mathfrak{u}}}
               \frac{({\overline{\mathcal{Z}}}^{(3)}_{q}
               ({\mathfrak{u}}))^{1-q} -1}
              {1-q} 
              \equiv  - k {\mathfrak{u}} \; \frac{\partial}
                {\partial {\mathfrak{u}}}
               \frac{{\mathfrak{c}}^{(3)}-1}
              {1-q}, 
\label{C3_Z3}
\eeq
even though an explicit evaluation of the internal energy $U_{q}^{(3)}$ by
integrating (\ref{U3_Z3}) may not be feasible.
The equivalence property in (\ref{C3_Z3}) is based on the equality (\ref{ur4}).
A perturbative calculation of the specific heat employing (\ref{C3_Z3}) 
is now accomplished  by using the sum $\mathfrak{c}^{(3)}$ of the 
$q$-weights in the third constraint picture. 
The connection formula (\ref{pro_link}) 
for the probabilities along with the transformation equation 
(\ref{uprime_u}) allows a systematic perturbative evaluation of the relevant 
quantity $({\overline{\mathcal{Z}}}^{(3)}_{q})^{1-q}$ appearing in the rhs 
of (\ref{C3_Z3}). A shortcoming of the method, however, 
is that it is imperative to evaluate the said quantity, say, at the 
order $(1 - q)^{3}$ for obtaining the specific heat, via (\ref{C3_Z3}), at 
the perturbative order $(1 - q)^{2}$. 

\par

Following the above description we now obtain a perturbative series for the 
generalized partition function raised to the exponent $(1 - q)$ by using 
the equations (\ref{pro_link}, \ref{ur4}, \ref{uprime_u}):  
\beq
[\overline{\mathcal{Z}}^{(3)}_q(\mathfrak{u})]^{1-q} 
= Z(\mathfrak{u})^{1-q} \displaystyle  \left(\sum_{n=0}^{\infty}(1-q)^n 
\sum_{\ell=0}^{2n-1} \mathfrak{z}_{n \ell}(\mathfrak{u}) 
(\mathcal{K}(\mathfrak{u}))^{\ell}\right).
\label{Gen_Z_Zcl}
\eeq
The first few coefficients $\mathfrak{z}_{n \ell}(\mathfrak{u})$ in the above 
series read as follows: 
\begin{eqnarray} 
\mathfrak{z}_{00}(\mathfrak{u}) &=& 1, \qquad \qquad
\mathfrak{z}_{10}(u) = N(3-\mathfrak{u}), \qquad \qquad
\mathfrak{z}_{11}(\mathfrak{u}) = N,   \nonumber \\
\mathfrak{z}_{20}(\mathfrak{u}) &=&  \frac{3}{2} (9N-1)N - 6N^2 \mathfrak{u} 
+ \frac{1}{2} (4+7N) N {\mathfrak{u}}^2 - N^2 {\mathfrak{u}}^3 + N(3 
+ {\mathfrak{u}}^2) \ln f(\mathfrak{u}), \nonumber \\ 
\mathfrak{z}_{21}(\mathfrak{u}) &=& -\frac{3}{2}(3+2N)N
+2N^2{\mathfrak{u}}+(1 + N)N {\mathfrak{u}}^{2} - 3 N  
\ln f(\mathfrak{u}),  \nonumber \\ 
\mathfrak{z}_{22}(\mathfrak{u}) &=& -\frac{1}{2}(10+11N)N
+ N^2 {\mathfrak{u}} - N \ln f(\mathfrak{u}), \qquad  
\mathfrak{z}_{23}(\mathfrak{u}) = -(1 + N)N, \nonumber \\
\mathfrak{z}_{30}(\mathfrak{u}) &=& \frac{1}{4}(1-27N+234N^2) N
+\frac{3}{2}(1-21N) N^2 \mathfrak{u} 
-\frac{1}{8}(33+168N-72N^2)N{\mathfrak{u}}^2\nonumber \\ 
& & +\frac{1}{6}(39+17N) N^2{\mathfrak{u}}^3 
+ \frac{1}{12} (53 + 33 N + 18 N^2) N {\mathfrak{u}}^4
-(1 + N)N^{2}{\mathfrak{u}}^5 \nonumber \\ & & 
+( 3N^2(6-\mathfrak{u})-\frac{1}{2} (17+12N) N 
{\mathfrak{u}}^2+4N^2\mathfrak{u}^3
+(1 + N)N \mathfrak{u}^4)\ln f(\mathfrak{u}) \nonumber \\ 
& & - \frac{5}{2} N \mathfrak{u}^2 (\ln f(\mathfrak{u}))^2, \nn\\ 
\mathfrak{z}_{31}(\mathfrak{u})&=& \frac{1}{4}(33+174N+18N^2)N
-\frac{9}{2}(3+2N) N^2{\mathfrak{u}}-\frac{1}{12}(445+600N+318N^2)
N{\mathfrak{u}}^{2} \nonumber \\ 
& & +15(1 + N)N^2{\mathfrak{u}}^{3}+(N+2)N{\mathfrak{u}}^{4} 
+\Big( 6(3+2N)N-9 N^2 \mathfrak{u}-16 (1 + N)N \mathfrak{u}^{2} \nn\\ 
& & + 2N^2\mathfrak{u}^{3} \Big) \ln f(\mathfrak{u})  
+ (6-\mathfrak{u}^2) N (\ln f(\mathfrak{u}))^2, \nonumber \\
\mathfrak{z}_{32}(\mathfrak{u}) &=& \frac{1}{8}(547+1050 N + 432 N^2)N 
-\frac{1}{2} (79+73N)N^2{\mathfrak{u}}\nn\\
& &-\frac{1}{6}(169+159N+90N^2) N {\mathfrak{u}}^{2} + 4(1 + N)N^2 
{\mathfrak{u}}^{3}\nn\\ 
& &+ \Big(\frac{1}{2}(89+84N) N-10 N^2 \mathfrak{u}-4(1 + N)N 
\mathfrak{u}^{4}\Big) \ln f(\mathfrak{u}) 
+ \frac{11}{2}N (\ln f(\mathfrak{u}))^2, \nn\\
\mathfrak{z}_{33}(\mathfrak{u}) &=& \frac{1}{12}(907+1212N+722N^2)N
-21(1 + N) N^2 \mathfrak{u} - (5+4N+2N^2) N \mathfrak{u}^{2}  \nonumber \\ 
& & + (22(1 + N)N - 2 N^2\mathfrak{u})\ln f(\mathfrak{u}) 
+ N (\ln f(\mathfrak{u}))^2, \nonumber \\ 
\mathfrak{z}_{34}(\mathfrak{u}) &=&\frac{1}{4} (107 +119N+78N^2) N
-3(1 + N)N^2 \mathfrak{u} +3(1 + N)N \mathfrak{u} 
\ln f(\mathfrak{u}), \nonumber \\ 
\mathfrak{z}_{35}(\mathfrak{u}) &=& (3+3N+2N^{2})N.
\label{Z_coeff}
\end{eqnarray}
The alternate method of evaluation of the specific heat may now be completed 
by substituting the perturbative series (\ref{Gen_Z_Zcl}, \ref{Z_coeff}) in 
the defining relation (\ref{C3_Z3}). The results obtained in this 
parallel procedure completely agrees with our previous evaluation 
presented in (\ref{C3_expn}) and
(\ref{cprcr}). However, as we remarked earlier, an evaluation of the specific 
heat at the perturbative order $(1 - q)^{n}$ using the property (\ref{C3_Z3}) 
necessitates evaluation of the relevant sum $\mathfrak{c}^{(3)}$ of the 
$q$-weights at the succeeding order $(1 - q)^{n+1}$. Our prior evaluation of 
the specific heat based on direct determination (\ref{U3_series}) of the 
internal energy that follows from the thermodynamic transformation property 
(\ref{int_3_trans}) requires computation of \textit {all} pertinent quantities 
only up to the level $(1 - q)^{n}$, and, therefore, has an advantage in 
this respect.   

\par

We now obtain various physically relevant limits of the specific heat 
(\ref{C3_expn}). For this purpose we express the specific heat (\ref{C3_expn}) 
in terms of the variable $u$ introduced in (\ref{Z_1_K}), rather than the 
scaled quantity $\mathfrak{u}$ defined in (\ref{vol_scale}). The translation
of our results (\ref{C3_expn}, \ref{cprcr}) to the corresponding quantities 
expressed in terms of the variable $u$ may be most 
succinctly expressed by the functional form 
\beq
C^{(3)}_{q} (u, \ln Z(u)) = C^{(3)}_{q} (\mathfrak{u}, \ln f(\mathfrak{u})),
\label{C_tran}
\eeq     
where the classical partition function $Z(u)$ is given in (\ref{Z_factor}). 
The lhs of (\ref{C_tran}) is easily obtained by replacing the polynomials in 
$\mathfrak{u}$ in the coefficients (\ref{cprcr}) by identical polynomials in 
the variable $u$, and simultaneous substitution of $\ln f(\mathfrak{u})$ 
factors therein with the corresponding quantities in the variable 
$\ln Z(u)$. For the sake 
of brevity we do not reproduce the full expression of the lhs 
in (\ref{C_tran}).

\par

({\sf i}) For the nonrelativistic gas the limiting value is given by $u \gg 1$.
We substitute the asymptotic expansion for the ratio of Bessel functions 
given in the Appendix (\ref{cprbsa}) in the expression of the 
specific heat obtained via the replacement (\ref{C_tran}) in the equations 
(\ref{C3_expn}, \ref{cprcr}). We also note that in the $u \gg 1$ regime the
classical relativistic partition function defined in (\ref{part_rel}) reduces 
[\cite{GNS95}] to its nonrelativistic analog $Z_{NR}(T, V, N)$:
\beq
Z (T, V, N)|_{u \gg 1} = Z_{NR}(T, V, N) \equiv \frac{V^{N}}{N!}\;
\left(\frac{2 \pi m k T}{h^{2}}\right)^{\frac{3N}{2}},
\label{Z_ZNR}
\eeq
where we have used the dimension $D = 3$. With the above facts in mind, we now 
obtain the nonrelativistic limit of the perturbative series for the specific 
heat of the ideal gas in the third constraint picture as 
\bea
\frac{2}{3}\frac{C^{(3)}_{q}\Big{|}_{NR}}{Nk} &=& 1+(1-q)\;(3 N + 
\ln Z_{NR})\nn\\ 
& &- \frac{(1-q)^{2}}
{8}\Big(3N(2 - 21N) - 36 N \ln Z_{NR} - 4 (\ln Z_{NR})^{2}\Big)+\ldots . 
\label{cprcnr}
\eea
The above series reproduces the specific heat of the nonrelativistic ideal 
gas [\cite{SA}] in the neighborhood of $q \rightarrow 1$.

\par

({\sf ii}) Another interesting limit is the extreme relativistic case that has 
been discussed in detail in Sec. \ref{ultra}, where we have obtained the
corresponding specific heat (\ref{i9}) in a closed form. Here, following the 
transition to the variable $u$ that has been explained in the context of 
(\ref{C_tran}), we consider the massless $m \rightarrow 0$ limit of the 
perturbative series for the specific heat (\ref{C3_expn}, \ref{cprcr}). In 
the $m \rightarrow 0$ limit we observe that the classical partition function
(\ref{part_rel}) reduces to
\beq
Z (T, V, N)|_{u \rightarrow 0} = Z_{m = 0}(T, V, N),
\label{Z_ZNcl_m0}
\eeq  
where the classical partition function $Z_{m = 0}(T, V, N)$ in the massless 
case is given in (\ref{cl_m0_Z}). The  $u\rightarrow0$ limiting value of the 
specific heat obtained from (\ref{C3_expn}, \ref{cprcr}) reads 
\bea
\frac{C_{q}^{(3)}\Big{|}_{u \rightarrow 0}}{3Nk} &=& 1 + (1-q)(6 N + 
\ln Z|_{m = 0})\nn\\ 
& &- \frac{(1-q)^{2}}{2}\Big(3 N (1 - 21 N)
- 18 N \ln Z|_{m = 0} + (\ln Z|_{m = 0})^{2}\Big)+\ldots.
\label{cprcer}
\eea
As a further check on our results (\ref{C3_expn}, \ref{cprcr}) the limiting 
series (\ref{cprcer}) in the vicinity of $q \rightarrow 1$ completely agrees 
with a perturbative expansion of the exact value (\ref{i9}) of the 
specific heat for the extreme relativistic ideal gas for the dimension $D = 3$.

%
%
%
%
%
%
\setcounter{equation}{0}
\section {Remarks }
\label{remark}
We have considered a canonical ensemble of $N$ particles of a relativistic 
ideal gas, and found its specific heat in the third constraint scenario. In 
the extreme relativistic limit the generalized partition function, the 
internal energy, and, therefore, the specific heat may be exactly evaluated. 
This makes it possible for us to observe the nature of the 
singularities. The generalized partition function exhibits 
simple poles on the $q$-plane at $q=1+\frac{1}{n},\; n=1,2,\ldots,DN$, where 
the factor $D N$ equals the number of degrees of freedom. The canonical 
ensemble is, consequently, well-defined in the parametric range
$0 < q < 1 + \frac{1}{D N}$. As $N$ increases, the 
singularities approach the limit $q \rightarrow 1^{+}$ that represents 
extensive statistical mechanics. This agrees with the earlier observation 
[\cite{AK}] that a grand canonical ensemble of a nonextensive relativistic 
ideal gas does not exist in the $q > 1$ regime. We also notice that in the 
present case of relativistic gas the thermodynamic limit and the extensive 
limit do not commute. For other systems similar results also hold 
[\cite{SA},\cite{RCN}]. 

\par

For the general case of a relativistic ideal gas with arbitrary mass of 
the molecules we used a perturbative mechanism developed in [\cite{RCN}]. The 
specific heat was obtained as a perturbative series 
in the nonextensivity parameter $(1-q)$ 
up to the second order. The evaluation was performed by using two different 
methods: a procedure that directly links the internal energies in the second 
and third constraint pictures via a transformation, and the traditional 
generalized partition function based approach. To our knowledge the former 
method has not been used earlier. The second approach requires the 
evaluation of the sum of the $q$-weights $\mathfrak{c}^{(3)}$ defined in 
(\ref{ur4}) up to a perturbative order higher than the prescribed order 
of the specific heat, whereas in the first method 
based on the transformation of internal energies all the quantities are 
uniformly computed up to the required perturbative order of evaluation of the 
specific heat.  Both procedures generate identical perturbative series for the 
specific heat. As a bonus, the former approach easily produces the series 
for the internal energy in the third constraint picture. The known series of 
internal energy, in turn, produce the specific heat. The 
nonrelativistic and the extreme relativistic limits of the 
said perturbative series of the specific heat agree with known respective 
expressions.

\par

Using the integral representations of the gamma function, Prato [\cite{DP}] 
observed that in the context of the second constraint picture, it is possible 
to connect the generalized partition function for a nonextensive statistical 
system with the partition function of the corresponding extensive 
$(q = 1)$ system for both $q > 1$ 
and $q < 1$ domains. We now demonstrate such relations for the extreme 
relativistic ideal gas that has been exactly solved in our work. Moreover, 
such integral representations provide alternate derivation of the 
perturbative expansion scheme for the generalized partition function of a 
relativistic ideal gas with particles of arbitrary mass. The Hilhorst integral 
for $q>1$ region reads  
\begin{equation}
\mathcal{Z}^{(2)}(\beta,V,N)= \displaystyle \int_{0}^{\infty} dt\; 
\mathfrak{G}(t)\; Z(t \beta,V,N),
\label{q_G_1}
\end{equation}
where the kernel $\mathfrak{G}(t)$ is given by [\cite{DP}]
\begin{equation}
\mathfrak{G}(t) = \frac{1}{{(q-1)}^{\frac{1}{q-1}}\;\Gamma(\frac{1}{q-1})} \;\; 
t^{\frac{1}{q-1}-1} \;\;\exp \Big(- \frac{t}{q-1}\Big).
\label{ker_qG1}
\end{equation}
The classical partition function $Z_{m=0}(\beta,V,N)$ for extreme relativistic 
ideal gas is given in (\ref{cl_m0_Z}). The corresponding nonextensive 
partition function in the second constraint picture is obtained via 
(\ref{q_G_1}) for the region $q>1$:
\begin{eqnarray}
\mathcal{Z}^{(2)}(\beta,V,N)\Big|_{m = 0} &=& Z_{m = 0}(\beta,V,N)
\left(\frac{\Gamma\Big(\frac{1}{q-1}-DN\Big)}{(q-1)^{DN} \Gamma(\frac{1}{q-1})} 
\right)\nn\\
&=& Z_{m = 0}(\beta,V,N)\;\;\displaystyle\prod_{n = 1}^{D N}\;
\frac{1}{1 + (1 - q) \,n}. 
\label{Z2_m0_G1}
\end{eqnarray}
For an arbitrary mass of the molecules we consider the kernel (\ref{ker_qG1}) 
perturbatively. In the limit $q \rightarrow 1^{+}$, it behaves as a 
distribution comprising of delta function and its derivatives having support 
at $t = 1$:
\begin{equation}
\mathfrak{G}(t) = \delta(t-1) + \frac{1}{2} \;(q - 1)\;\; \frac{{\partial}^2}
{\partial t^2} \delta(t-1) - \frac{1}{3} \;(q-1)^2\;\;
\left(\frac{{\partial}^3}{\partial
t^3}  - \frac{3}{8} \frac{{\partial}^4}{\partial t^4}\right) \delta(t-1) 
+ \ldots.
\label{K_Dis}
\end{equation}
Substituting the above series of distributions in the Hilhorst integral 
(\ref{q_G_1}), we derive a perturbative connection formula between the 
generalized partition function in the second constraint scenario and the 
classical Boltzmann-Gibbs partition function. This connection formula agrees
precisely with  (\ref{part_rel}) obtained by using an infinite product 
expansion of the $q$-exponential.

\par

For the complementary $q<1$ region, the generalized partition function in the 
second constraint framework is expressed [\cite{DP}]  in terms of the 
Boltzmann-Gibbs partition function as a contour integral on a complex plane:
\begin{equation}
\mathcal{Z}^{(2)}(\beta,V,N)= \frac{i}{2 \pi} \displaystyle 
\oint_{\mathsf C} dt \; \overline{\mathfrak{G}}(t)\; Z(t \beta,V,N),
\label{q_L_1}
\end{equation}
where the kernel reads
\begin{equation}
\overline{\mathfrak{G}}(t) = \frac{\Gamma(\frac{2-q}{1-q})}{(q-1)^{-(\frac{2-q}
{1-q})+1}}\;\; (-t)^{-\frac{2-q}{1-q}}\;\; \exp\Big(- \frac{t}{q-1}\Big).
\label{ker_qL1} 
\end{equation}
The contour $\mathsf {C}$ on the complex plane in (\ref{q_L_1}) is comprised 
[\cite{DP}] of the segments $\displaystyle\{(\infty, \varepsilon), 
(t = \varepsilon\, \exp (i\,\vartheta),
0 < \vartheta < 2 \pi), (\varepsilon\, \exp(i\, 2\pi^{-}),\;
\infty\;\exp(i\, 2\pi^{-}))|\, \varepsilon \rightarrow 0\}$.
Employing the kernel (\ref{ker_qL1}) as before, we compute the generalized 
partition function for the extreme relativistic ideal gas in the $q < 1$ 
domain:    
\bea
\mathcal{Z}^{(2)}(\beta,V,N)\Big{|}_{m = 0} &=& Z_{m = 0}(\beta,V,N)
\left(\frac{\Gamma\Big(\frac{2-q}{1-q}\Big)}{(1-q)^{DN} 
\Gamma\Big(\frac{2-q}{1-q}+DN\Big)}\right)\nn\\
&=& Z_{m = 0}(\beta,V,N) \displaystyle \prod_{n = 1}^{D N}\;
\frac{1}{1 + (1 - q) \,n}. 
\label{Z2_m0_L1}
\eea
As expected, the expressions (\ref{Z2_m0_G1}) and (\ref{Z2_m0_L1}) produce 
identical results for the generalized partition function for the extreme
relativistic ideal gas signalling continuity at the extensive parametric value 
$q = 1$.

\bigskip

%
%
%
\setcounter{equation}{0}
\section*{APPENDIX}
\renewcommand{\theequation}{A.\arabic{equation}}

1. The translation of the argument of $\mathcal{K}(u)$ defined in 
(\ref{K_ratio}) is given by the following Taylor series:
\bea
\mathcal{K}(u) &=& \mathcal{K}(\mathfrak{u}) + (1-q) 
\Bigg((6N+\ln f(\mathfrak{u})){\mathfrak{u}}^{2} 
- 2N {\mathfrak{u}}^{3} -\Big(4(6N+\ln f(\mathfrak{u}))-8N\mathfrak{u}\nn \\ 
& & -2N{\mathfrak{u}}^{2}\Big)\mathcal{K}(\mathfrak{u}) 
- \Big((14N+\ln f(\mathfrak{u}))
-2N \mathfrak{u}\Big)(\mathcal{K}(\mathfrak{u}))^{2} 
- 2N (\mathcal{K}(\mathfrak{u}))^{3} \Bigg) \nn \\
& & + (1-q)^{2} \Bigg(-3\Big(\frac{1}{2}(1+69N+22\ln f(\mathfrak{u})) N 
+ (\ln f(\mathfrak{u}))^{2}\Big){\mathfrak{u}}^{2} + (75N \nn \\ 
& & +13 \ln f(\mathfrak{u}))N {\mathfrak{u}}^{3}+ \frac{1}{2}
(9-15N+4N \ln f(\mathfrak{u}))N {\mathfrak{u}}^{4} 
- 2 N^{2} {\mathfrak{u}}^{5}+\Big(6(1+45N \nn \\ 
& & +14 \ln f(\mathfrak{u}))N + 8 (\ln f(\mathfrak{u}))^{2} -12 (17N+3 
\ln f(\mathfrak{u})) N \mathfrak{u} 
- \frac{1}{2} (57N + 230N^{2} \nn \\
& &  + 78 N \ln f(\mathfrak{u}) + 2 (\ln f(\mathfrak{u}))^{2}) 
{\mathfrak{u}}^{2} 
+ (65 N+4 \ln f(\mathfrak{u}))N {\mathfrak{u}}^{3}- 2N(N-1) {\mathfrak{u}}^{4}
\Big) \mathcal{K}(\mathfrak{u}) \nn \\
& & + \Big(\frac{1}{2}(87+975N+258 \ln f(\mathfrak{u})) N 
+ 6 (\ln f(\mathfrak{u}))^{2} 
- (247N+25 \ln f(\mathfrak{u})) N \mathfrak{u} \nn \\
& & - (23 + 38N + 8 \ln f(\mathfrak{u})) N {\mathfrak{u}}^{2} 
+ 12 N^{2} {\mathfrak{u}}^{3}\Big) {\mathcal{K}(\mathfrak{u})}^{2} 
+ \Big(\frac{1}{2}(105+574N \nn \\
& & + 102 \ln f(\mathfrak{u})) N+ (\ln f(\mathfrak{u}))^{2} 
- (89N+4 \ln f(\mathfrak{u})) N \mathfrak{u} 
- 4 (1 + N) N {\mathfrak{u}}^{2} \Big) (\mathcal{K}(\mathfrak{u}))^{3} \nn \\
& & +\Big(\frac{1}{2}(37+139N + 12 \ln f(\mathfrak{u})) N - 10 N^{2} 
\mathfrak{u}\Big) (\mathcal{K}(\mathfrak{u}))^{4} \nn \\
& & + 2(1+3N)N(\mathcal{K}(\mathfrak{u}))^{5}\Bigg) + \ldots.
\label{bess_sr_u2w}
\eea
2. The asymptotic expansion of the ratio of Bessel functions in the 
$u \gg 1$ region reads 
\beq
\frac{K_{1}(u)}{K_{2}(u)} \approx 1 - \frac{3}{2u} 
+ \frac{15}{8u^{2}} - \frac{15}{8u^{3}}
 + \frac{135}{128u^{4}} + \frac{45}{32u^{5}} - \frac{7425}{1024 u^{6}} 
+ \frac{375}{32 u^{7}} - \frac{1095525}{32768u^{8}} + \ldots.
\label{cprbsa}
\eeq

%
%
%
\section*{Acknowledgements} 
R. Chandrashekar and S.S. Naina Mohammed would like to acknowledge the 
fellowships received from the Council of Scientific and Industrial 
Research(India) and the University Grants Commission (India), respectively.
%

%
%
%

\end{document}